\documentclass[%
 reprint,
superscriptaddress,
 amsmath,amssymb,
 aps,
prl,
]{revtex4-1}

\usepackage{graphicx}
\usepackage{svg}
\usepackage{siunitx}
\usepackage{acro}

\DeclareSIUnit\angstrom{\text{Å}}

\DeclareAcronym{FWHM}{
  short = FWHM,
  long  = full width at half maximum
}
\DeclareAcronym{PIC}{
  short = PIC,
  long  = photonic integrated circuit
}
\DeclareAcronym{SOI}{
  short = SOI,
  long  = silicon on insulator
}
\DeclareAcronym{BOX}{
  short = BOX,
  long  = burried oxide
}
\DeclareAcronym{FOM}{
  short = FOM,
  long  = figure of merit
}
\DeclareAcronym{TE}{
  short = TE,
  long  = transverse electric
}
\DeclareAcronym{TM}{
  short = TM,
  long  = transverse magnetic
}
\DeclareAcronym{GC}{
  short = GC,
  long  = grating coupler
}
\DeclareAcronym{RIE}{
  short = RIE,
  long  = reactive ion etching
}
\DeclareAcronym{SEM}{
  short = SEM,
  long  = scanning electron microscope
}
\DeclareAcronym{RBF}{
  short = RBF,
  long  = radial basis function
}

\usepackage[colorlinks=true, linkcolor=blue, citecolor=blue, urlcolor=blue]{hyperref}
\usepackage{cleveref}
\crefname{figure}{Figure}{Figures}
\Crefname{figure}{Figure}{Figures}
\crefname{equation}{Equation}{Equations}
\Crefname{equation}{Equation}{Equations}
\usepackage{orcidlink}

\makeatletter
\renewcommand{\fnum@figure}{\textbf{Figure~\thefigure}}
\makeatother

\begin{document}


\title{Inverse-Designed Grating Couplers with Tunable Wavelength via Scaling and Biasing}

\author{Lorenz J. J. Sauerzopf \orcidlink{0009-0001-7652-5044}}
    \affiliation{TUM School of Computation, Information and Technology, Technical University of Munich, 80333 Munich, Germany}%
    \affiliation{Walter Schottky Institute, Technical University of Munich, 85748 Garching, Germany}%
    \affiliation{Munich Center for Quantum Science and Technology (MCQST), 80799 Munich, Germany}%
\author{Fabian Becker \orcidlink{0000-0002-4447-4211}}
    \affiliation{TUM School of Computation, Information and Technology, Technical University of Munich, 80333 Munich, Germany}%
    \affiliation{Walter Schottky Institute, Technical University of Munich, 85748 Garching, Germany}%
    \affiliation{Munich Center for Quantum Science and Technology (MCQST), 80799 Munich, Germany}%
\author{Kai M\"uller \orcidlink{0000-0002-4668-428X}}
    \affiliation{TUM School of Computation, Information and Technology, Technical University of Munich, 80333 Munich, Germany}%
    \affiliation{Walter Schottky Institute, Technical University of Munich, 85748 Garching, Germany}%
    \affiliation{Munich Center for Quantum Science and Technology (MCQST), 80799 Munich, Germany}%

\date{October 31, 2025}

\begin{abstract}
\begin{center}
    \includegraphics{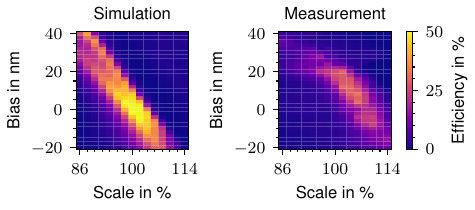}
\end{center}

\Aclp*{PIC} are heavily researched devices for telecommunication, biosensing, and quantum technologies. Wafer-scale fabrication and testing are crucial for reducing costs and enabling large-scale deployment. \Aclp*{GC} allow non-invasive measurements before packaging, but classical designs rely on long tapers and narrow bandwidths. In this work, we present compact, inverse-designed \aclp*{GC} with broadband transmission. We optimized and fabricated arrays of devices and characterized them with a 4f-scanning setup. The nominal design reached simulated efficiencies of \SI{52}{\percent}, while measurements confirmed robust performance with up to \SI{32}{\percent} efficiency at the target \SI{1540}{\nano\meter} wavelength and \SI{46}{\percent} at shifted wavelengths. Without scaling and contour biasing, the measured efficiency at the target wavelength drops to only \SI{4.4}{\percent}. Thus, a key finding is that systematic scaling and edge biasing recover up to an eightfold improvement in efficiency. These inverse-designed \aclp*{GC} can be efficiently corrected post-design, enabling reliable performance despite fabrication deviations. This approach allows simple layout adjustments to compensate for process-induced variations, supporting wafer-scale testing, cryogenic photonic applications, and rapid design wavelength tuning.
\end{abstract}

\acresetall

\maketitle

\section{Introduction} \label{Introduction}
\Acp{PIC} play a central role in current research~\cite{butt_lighting_2025,terrasanta_photonic_2025} and industrial applications such as transceivers~\cite{qian_hybrid_2025} and optical multiplexing~\cite{jiang_terabitpersecond_2025}.
The \ac{SOI} platform with a \SI{220}{\nano\meter} device layer has become the industry standard~\cite{dan-xia_xu_silicon_2014} for telecom C-band communication.
High-efficiency coupling is essential to improve the power performance of such systems. In-plane trident fiber edge-couplers reach efficiencies of up to \SI{81}{\percent}~\cite{hatori_trident_2014}, but a \SI{3}{\micro\meter} misalignment already causes losses greater than \SI{12}{\decibel}~\cite{liu_high_efficiency_2025}. Suspended edge waveguides on \ac{SOI} achieve up to \SI{65}{\percent} coupling efficiency~\cite{Fang_suspended_2010,Galan_suspended_2007}.
Both methods provide coupling \SI{3}{\decibel} bandwidths around \SI{100}{\nano\meter}~\cite{marchetti_coupling_2019}.

Cryogenic applications, however, make fiber edge-coupling difficult due to space restrictions and fiber misalignment during cooldown. \Acp{GC} overcome these limitations with free-space coupling. They also enable wafer-scale testing during fabrication, which is valuable for industry. Conventional \acp{GC}, however, require angled incidence or advanced fabrication techniques such as bottom mirrors or top wedges to reach high efficiencies comparable to fiber coupling~\cite{marchetti_coupling_2019}.

Conventional \acp{GC} rely on long tapered waveguides, often hundreds of \si{\micro\meter} in length, and remain limited to narrow bandwidths of about \SI{30}{\nano\meter}~\cite{marchetti_high-efficiency_2017}. Inverse design breaks this limitation by enabling compact, small-footprint \acp{GC}~\cite{kim_inverse_2025}. Using adjoint-based topology optimization, these devices emerge from direct solutions of Maxwell’s equations, revealing pixelated designs inaccessible to brute-force search~\cite{georgieva_feasible_2002,veronis_method_2004,jelena_specialissueinvite_2023}.

Inverse-designed \acp{GC} have demonstrated excellent performance in diamond platforms, with strong agreement between simulated and measured wavelengths~\cite{dory_inverse_designed_2019}. Nevertheless, unintentionally wavelength shifts in fabricated \acp{GC} have not yet been fully eliminated~\cite{marchetti_high-efficiency_2017}, especially in complex material stacks and device designs~\cite{hammond_multi_layer_2022}.
This wavelength discrepancy is also a long-standing issue in photonic crystals, where even tiny changes in defect spacing shift the transmission wavelength~\cite{foresi_photonic-bandgap_1997,joannopoulos_photonic_2011,krauss_waveguide_1997,krauss_photonic_1999}.
The root cause lies in fabrication variations -- most notably positive and negative contour biasing -- which introduce spectral shifts in both photonic crystals~\cite{abulnaga_design_2025} and inverse-designed devices.

This work introduces a method for robust \acp{GC} in cryogenic telecom applications, where vertical coupling, small footprint, and fabrication tolerance are critical. We designed and fabricated inverse-designed \acp{GC} and characterized the scaled and contour biased devices with a 4f-scanning setup. The results show that these adjustments provide a powerful route to fine-tune the wavelength response and recover possible efficiency losses to due fabrication errors.

\section{Methods}\label{Methods}
We optimized the \ac{GC} with Lumerical FDTD and a modified version of lumopt and its topology optimizer~\cite{lumopt}. The simulated device stack consisted of a \SI{216}{\nano\meter} Si layer ($n_{Si}=3.5$) on a \SI{3}{\micro\meter} SiO \ac{BOX} substrate ($n_{SiO}=1.78$). The algorithm could freely vary the refractive index per pixel to create the structure, within a \qtyproduct{5 x 5}{\micro\meter} region discretized with a \SI{20}{\nano\meter} mesh. A \SI{550}{\nano\meter}-wide waveguide is connected to the region edge, while a Gaussian source illuminated the region center normal to the surface. To enforce broadband performance, we simulated 11 wavelengths from \SI{1520}{\nano\meter} to \SI{1570}{\nano\meter}.

We defined the figure of merit (FOM) as the summed overlap between forward-coupled light and the fundamental \ac{TE} mode of the waveguide, measured at a mode field monitor placed at a distance of \SI{1.8}{\micro\meter} from the optimization region. To prevent non-fabricable features, we applied a low-pass filter with \SI{80}{\nano\meter} radius in the final optimization steps, smoothing sharp edges and eliminating small artifacts. The design was represented as a pixelated refractive index map with independently variable pixels.

\begin{figure}[t!]
    \includesvg[width=0.5\linewidth]{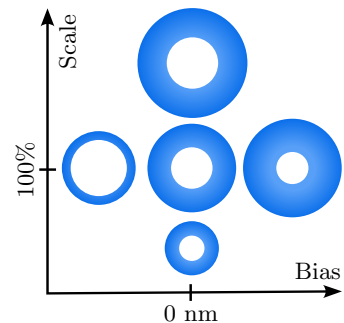}
    \caption{Illustration of scaling and biasing applied to a simple donut structure. Scaling preserves the design proportions, while biasing shifts the contour inward or outward to emulate fabrication over- and underetching.}
    \label{fig:effect}
\end{figure}

\begin{figure}[t!]
  \centering

    \begin{minipage}[t]{0.46\textwidth}
      \centering
      \makebox[0pt][l]{\hspace{-12.5em}\textbf{(a)}}
      \includesvg[width=\linewidth]{Figures/GC_waveguide}
    \end{minipage}

  \vspace{1.9em} 

  \begin{minipage}[b]{0.2\textwidth}
    \makebox[0pt][l]{\raisebox{0.68\linewidth}[0pt][0pt]{\hspace{-1.5em}\textbf{(b)}}}%
    \includegraphics[height=0.65\linewidth]{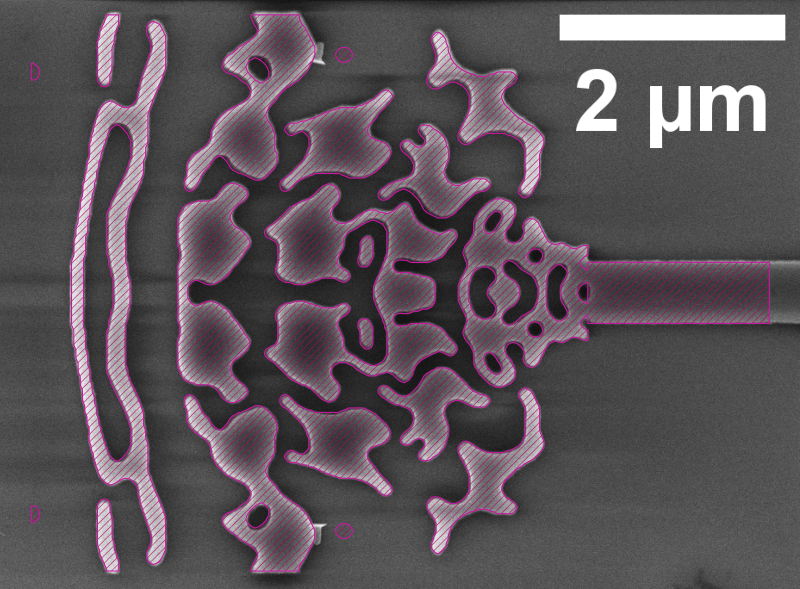}
  \end{minipage}
  \hspace{0.02\linewidth}
  \begin{minipage}[b]{0.2\textwidth}
    \makebox[0pt][l]{\raisebox{0.68\linewidth}[0pt][0pt]{\hspace{-1.5em}\textbf{(c)}}}%
    \includegraphics[height=0.65\linewidth]{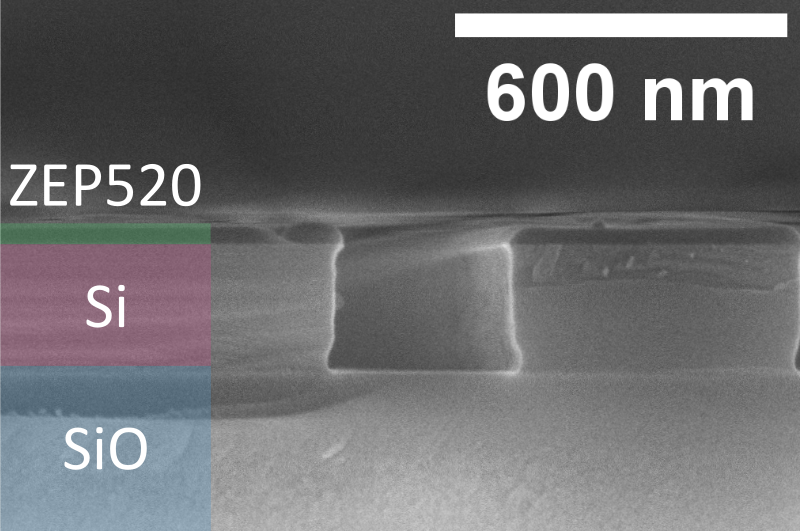}
  \end{minipage}

  \caption{\textbf{(a)} Photonic test device consisting of two \acp{GC} connected by a \SI{100}{\micro\meter}-long waveguide for coupling efficiency characterization. \textbf{(b)} \acs*{SEM} image of the nominal design with GDSII mask overlay. Small detached islands are visible but have negligible impact since the optical field vanishes at the coupler edges. \textbf{(c)} Cleaved sidewall showing footing at the Si–SiO\textsubscript{2} interface caused by the \ac{RIE} step.}
  \label{fig:combined}
\end{figure}

\Cref{fig:effect} highlights the role of scaling, which preserves proportions, and edge biasing, which shifts boundaries to emulate the effects of fabrication over- and underetching. \Cref{fig:combined}\textbf{a} shows the GDSII layout of the fabricated test structure. It consists of two \acp{GC} linked by a \SI{100}{\micro\meter} waveguide, one for input and one for output.

We fabricated arrays of devices with scale factors from \SI{86}{\percent} to \SI{114}{\percent} and biases from \SI{-20}{\nano\meter} to \SI{40}{\nano\meter}, in \SI{2}{\percent} and \SI{2}{\nano\meter} steps respectively. To prepare the devices for fabrication, we exported the optimized design to GDSII and post-processed it in Python using a modified version of gdshelpers~\cite{gehring_python_2019}. This workflow allowed us to apply scaling and edge biasing to arbitrary layouts. The layouts were processed with BEAMER (GenISys) to generate machine code for an EBPG 5150 electron beam writer. A diluted ZEP520A resist with a thickness of \SI{85}{\nano\meter} was used for pattern transfer. The exposure was done using a \SI{260}{\micro\coulomb\per\square\centi\meter} dose, \SI{100}{\kilo\volt} accelartion voltage, and \SI{10}{\nano\ampere} beam current. After developing in ZED N50 for \SI{65}{\second} and \SI{65}{\second} rinsing in isopropanol, both at \SI{5}{\celsius}, we etched the Si layer by cryogenic \ac{RIE} and removed the resist using ZDMAC.

After fabrication, the quality of the structure was assessed with \ac{SEM} imaging. As shown in \cref{fig:combined}\textbf{b}, the nominal design with \SI{100}{\percent} scale and \SI{0}{\nano\meter} bias aligns well with the GDSII layout, although small islands detached during processing. Their influence is negligible since the optical field strength is minimal at the GC edges. A cleaved sidewall image of a sample processed within the same \ac{RIE} run in \cref{fig:combined}\textbf{c} reveals footing at the Si–SiO\textsubscript{2} interface and serration at the Si–resist boundary, both caused by excessive etching.

Subsequently, we characterized the devices using a 4f-scanning mapper setup that enabled independent positioning of excitation and detection spots in the same focal plane on the \acp{GC}. Motorized mirrors optimized the alignment in-plane, while z-alignment for focusing was done manually. Polarization optics in excitation and detection ensured alignment with the \ac{TE} mode supported by the waveguide. A pellicle in the imaging path introduced polarization dependence, which we corrected for by referencing the measured device spectra to a silver mirror, assuming 100\% reflectivity.

For broadband excitation, we used a SuperK EVO EUL-10 white-light laser (NKT) and detected the output with a Kymera 193i spectrometer (Oxford Instruments). Each spectrum was background-subtracted, and an optimization algorithm adjusted excitation and detection spots iteratively to maximize the detected counts. To extract the device efficiency, we assumed a lossless waveguide and equal coupling efficiency in both in- and outcoupling. Under these conditions, the \ac{GC} efficiency $\eta_{GC}(\lambda)$ was determined as

\begin{equation}
    \eta_{GC}(\lambda) =  \sqrt{\frac{n_{raw}(\lambda) - n_{back}(\lambda)}{n_{mirror}(\lambda) - n_{back}(\lambda)} },
    \label{eq:gc_eff}
\end{equation}

with $n_{raw}(\lambda)$ the device counts, $n_{back}(\lambda)$ the background counts, and $n_{mirror}(\lambda)$ the silver-mirror reference. This procedure was repeated for all devices discussed in the Results section.

\section{Results}\label{results}
The lumopt optimization converged after 473 iterations, producing a design with a simulated peak efficiency of \SI{57}{\percent} at \SI{1540}{\nano\meter} for the nominal design with \SI{100}{\percent} scale and \SI{0}{\nano\meter} edge bias. When reimported as GDSII generated for fabrication, the efficiency dropped slightly to \SI{52}{\percent}, reflecting post-processing and discretization effects.

\begin{figure*}[t!]
    \centering
    \includegraphics[width=0.85\textwidth]{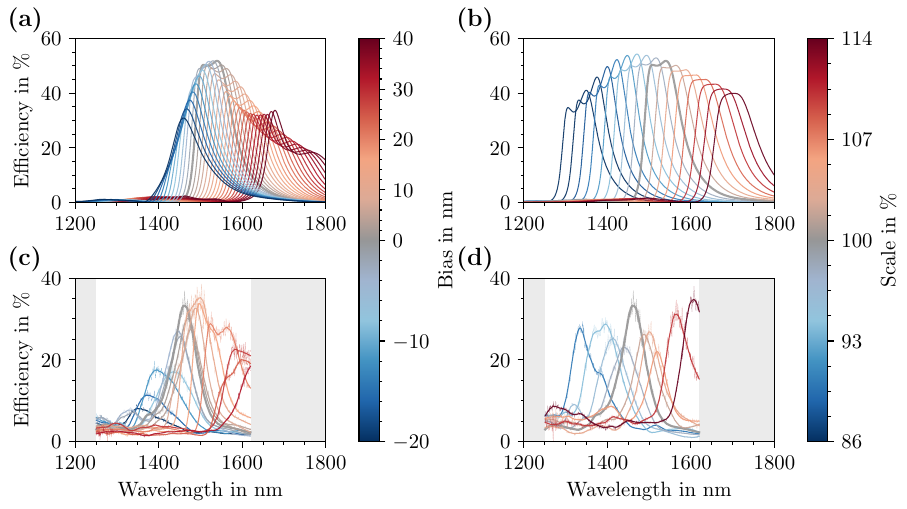}
    \caption{Simulated and measured efficiency spectra of \acp{GC}. \textbf{(a)} Simulation for fixed scale (\SI{100}{\percent}) and varying edge bias. \textbf{(b)} Simulation for fixed edge bias (\SI{0}{\nano\meter}) and varying scale. \textbf{(c)} and \textbf{(d)} Measured spectra from the 4f-scanning setup for the same parameters, shown as dotted raw data and solid smoothed curves. Thick gray spectra indicate the nominal design with \SI{0}{\nano\meter} edge bias or \SI{100}{\percent} scale. The gray area indicate where the measured data was truncated.}
    \label{fig:raw_data}
\end{figure*}

\Cref{fig:raw_data}\textbf{a} shows simulated efficiency spectra for different edge biases with a fixed scale of \SI{100}{\percent}. Negative biases shift the peak to shorter wavelengths, while positive biases shift it to longer wavelengths. A similar trend occurs for scaling in \cref{fig:raw_data}\textbf{b} with a fixed bias of \SI{0}{\nano\meter} for smaller and bigger scale respectively.
Measurements in \cref{fig:raw_data}\textbf{c-d} confirm these shifts experimentally: dotted lines show raw data, and solid lines are Savitzky–Golay–smoothed spectra used for reliable \ac{FWHM} extraction.
The nominal design is highlighted in in \cref{fig:raw_data} as thicker solid gray line and peaks at \SI{1462}{\nano\meter} with \SI{37}{\percent} efficiency.

\begin{figure*}[t!]
    \centering
    \includegraphics[width=1.0\textwidth]{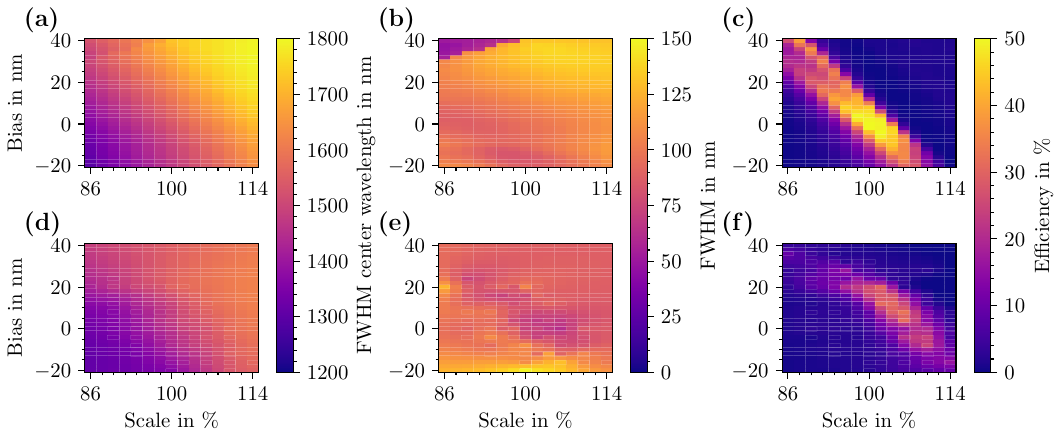}
    \caption{Simulation \textbf{(a–c)} and measurement \textbf{(d–f)} results of scaled and biased \acp{GC}. White cells mark the experimentally tested devices, while the rest are obtained by linear inter- or extrapolation. \textbf{(a)} and \textbf{(d)} Center wavelength extracted from the \ac{FWHM} of the efficiency spectrum. \textbf{(b)} and \textbf{(e)} \Ac{FWHM} bandwidth of the efficiency spectra. \textbf{(c)} and \textbf{(f)} Efficiency at the design wavelength of \SI{1540}{\nano\meter}.}
    \label{fig:data}
\end{figure*}

\Cref{fig:data} illustrates the influence of scaling and biasing on the device performance, comparing simulation (\cref{fig:data}\textbf{a–c}) and experiment (\cref{fig:data}\textbf{d–f}). In the measurement plots, white boxes mark the directly measured devices, while other data points are filled in using a linear-kernel \ac{RBF} interpolator. Both in simulation (\cref{fig:data}\textbf{a}) and experiment (\cref{fig:data}\textbf{d}), larger scales and positive biases consistently shift the spectrum toward longer wavelengths. Here, the reported wavelength corresponds to the \ac{FWHM} center rather than the spectral peak. Because the InGaAs detector loses sensitivity above \SI{1620}{\nano\meter}, peaks beyond this cutoff could not be resolved.

\Cref{fig:data}\textbf{b,e} compare the \ac{FWHM} bandwidths. The measured (\cref{fig:data}\textbf{e}) values are consistently narrower than those predicted (\cref{fig:data}\textbf{b}), indicating reduced bandwidth in practice. Importantly, the \ac{FWHM} was only evaluated for spectra where both sides of the bandwidth fell within the measurement window of \cref{fig:raw_data}. Spectra extending beyond the window were excluded. The apparent jump in \cref{fig:data}\textbf{b} at high bias values results from a transition between single- and double-peaked spectra, which produces an artificial drop in the calculated \ac{FWHM}.

Finally, \cref{fig:data}\textbf{c,f} show the efficiencies at the target wavelength of \SI{1540}{\nano\meter}. In simulation (\cref{fig:data}\textbf{c}), the nominal design (\SI{100}{\percent} scale, \SI{0}{\nano\meter} bias) provides the maximum efficiency of \SI{52}{\percent}. The measurements (\cref{fig:data}\textbf{f}) follow the same general trend, but the optimum is shifted. Remarkably, the device with \SI{+8}{\nano\meter} bias and \SI{106}{\percent} scale reaches \SI{32}{\percent} efficiency with a \ac{FWHM} of \SI{77}{\nano\meter} at the target wavelength, while the efficiency of the nominal design only reaches \SI{4.4}{\percent}. 
Across all devices, the overall highest measured efficiency was \SI{46}{\percent}, obtained at a center wavelength of \SI{1432}{\nano\meter} and a \ac{FWHM} of \SI{75}{\nano\meter} for a device with a \SI{94}{\percent} scale and a \SI{16}{\nano\meter} bias.
This confirms that fabrication variations require compensation through layout pre-scaling and edge biasing to achieve optimal peak performance.

The results confirm that inverse-designed \acp{GC} can match the efficiency and bandwidth of conventional fiber-based coupling~\cite{marchetti_coupling_2019}, but with the added advantage of eliminating the need for time-consuming fiber alignment. At the same time, we find that peak performance is not inherent to the nominal design. Instead, systematic scaling and edge biasing are essential to compensate for fabrication variations.

\section{Conclusion}

The efficiencies presented here should be viewed as lower-bound estimates, since the calculation in \cref{eq:gc_eff} assumes waveguides without loss.
However, if we assume a conservative waveguide loss of \SI{10}{\decibel\per\centi\meter}, the resulting loss is about \SI{2.3}{\percent} for a \SI{100}{\micro\meter} waveguide. This value is within the measuring error due to misalignment.
Beyond waveguide propagation losses, device performance can be further enhanced by optimizing the \ac{RIE} process, which would reduce sidewall roughness and thereby could increase efficiency of the \acp{GC}.

The manual z-alignment can further introduce variations between measurements.
Moreover, the InGaAs detector imposed a spectral cutoff at \SI{1620}{\nano\meter}, restricting access to the broadband efficiency predicted in simulation. As a result, absolute efficiency values of the measured \acp{GC} cannot be directly compared.
Nevertheless, the consistent spectral shifts observed in both simulation and experiment confirm that scaling and biasing reliably counteract fabrication-induced variations.

Finally, maintaining high polarization extinction would enable the same setup to support not only spatially separated excitation and detection, but also avoid scattered light detection using cross-polarization measurements via rotating the \acp{GC} by \SI{90}{\degree} to each other. Such versatility illustrates how inverse-designed \acp{GC} can serve as powerful tools for both wafer-scale testing and cryogenic applications.

Taken together, these results establish scaling and edge biasing as practical post-design strategies to recover efficiency and bandwidth otherwise lost to fabrication variations. Beyond individual device performance, this approach provides a general framework for making inverse-designed nanophotonic components more robust and transferable across fabrication platforms.  

Looking forward, this method may also be applied to other inverse design devices like mode, wavelength or power splitters. By combining compactness, robustness, and flexibility, inverse-designed \acp{GC} not only match the performance of fiber-based coupling but also directly address the demands outlined in the introduction: wafer-scale testing, cryogenic operation, and broadband compatibility. In addition, it allows fast adaptation of photonic devices regarding the wavelength, reducing development time and costs. In this way, they offer a scalable pathway toward the deployment of next-generation photonic and quantum technologies.

\section{Acknowledgements}\label{S_Acknowledge}
We gratefully acknowledge support from the German Federal Ministry of Research, Technology and Space (BMFTR) via the funding program quantum technologies - from basic research to market (project QPIS, contract number 16K1SQ033 and project QPIC-1, contract number 13N15855) and project 6G-life, the Bavarian State Ministry for Science and Arts (StMWK) via projects NEQUS, the Bavarian Ministry of Economic Affairs (StMWi) via project 6GQT, as well as from the German Research Foundation (DFG) under Germany’s Excellence Strategy EXC-2111 (390814868) and projects PQET (INST 95/1654-1) and MQCL (INST 95/1720-1).

\end{document}